\documentclass[12pt]{iopart}
\usepackage{iopams}
\usepackage{epsf,graphics,graphicx}

\def\re{\mathrm{Re}}
\begin{document}
\title{Correlation induced non-Abelian quantum holonomies}
\author{Markus Johansson$^1$}
\ead{markus.johansson@kvac.uu.se}
\author{Marie Ericsson$^1$}
\ead{marie.ericsson@kvac.uu.se}
\author{Kuldip Singh$^2$}
\ead{sciks@nus.edu.sg}
\author{Erik Sj\"oqvist$^{1,2}$}
\ead{erik.sjoqvist@kvac.uu.se}
\author{Mark S. Williamson$^{2,3}$}
\ead{m.s.williamson04@gmail.com}
\address{$^1$Department of Quantum Chemistry, Uppsala University, Box 518,
Se-751 20 Uppsala, Sweden, EU}
\address{$^2$Centre for Quantum Technologies, National University of Singapore,
3 Science Drive 2, 117543 Singapore, Singapore}
\address{$^3$Erwin Schr\"{o}dinger International Institute for Mathematical Physics,
Boltzmanngasse 9, 1090 Wien, Austria, EU}
\date{\today}
\begin{abstract}
In the context of two-particle interferometry, we construct a parallel transport condition that is based on
the maximization of coincidence intensity with respect to local unitary operations on one of the subsystems.
The dependence on correlation is investigated and it is found that the holonomy group is generally non-Abelian,
but Abelian for uncorrelated systems. It is found that our framework contains the L\'{e}vay geometric phase
[2004 {\it J. Phys. A: Math. Gen.} {\bf 37} 1821] in the case of two-qubit systems undergoing local $SU(2)$ evolutions.
\end{abstract}

\submitto{\JPA}
\maketitle

\section{Introduction}

The theory of holonomies and geometric phases associated with evolutions of a quantum system is by now a
well developed subject. The initial work by Berry \cite{berrr} on the Abelian geometric phase of adiabatic
evolutions of non-degenerate states has been extended in many directions, to include non-adiabatic evolutions
\cite{aharanovanandan} as well as non-Abelian holonomies of sets of degenerate states \cite{wilzcekzee,anandan},
and to mixed states \cite{sjoqvist}. It was subsequently discovered that the quantum geometric phase had
an early counterpart in the geometric phase discovered by Pancharatnam in the context of classical optics
\cite{panca, berry1,ramaseshan}. The Pancharatnam construction has been generalized to the non-Abelian
case by utilizing subspaces \cite{mead,anandanpines,kulting,sjokultabe}. More recently holonomies that
bear a relation to correlations have been constructed in the context of multipartite and lattice systems
\cite{williamsonvedral,williamsonvedral2,williamson09, wootters}.

The idea of this paper is to develop the concept of geometric phases in another direction and use two-particle
interferometry to construct correlation induced non-Abelian holonomies in a way that does not depend on
degeneracy, but on the ability to divide the system into spatially separated subsystems. Instead of considering
parallel transport of subspaces, we consider the natural tensor product structure of a bipartite state, induced
by the spatial separation of the two subsystems, and local unitary operations to define the parallel transport.

The parallel transport condition
to be introduced is similar to that of the Pancharatnam construction \cite{panca, berry1,ramaseshan},
in which two states $\mid\!{A}\rangle$ and $\mid\!{B}\rangle$ of a quantum system are defined to be in-phase
if their scalar product $\langle{B}\!\mid\!{A}\rangle$ is a positive number. This condition can be implemented
in a Mach-Zehnder interferometric setup, where the spatial state of the system prior to the last beam splitter
is a coherent superposition of the two paths \cite{wash1, wash2}. If we let $\mid\!{A}\rangle$ and
$\mid\!{B}\rangle$ be the internal states corresponding to the output of respective paths of the interferometer,
then Pancharatnam parallelity is achieved by shifting a $U(1)$ phase in one of the paths, such that the
interference intensity is maximal.

In two-particle interferometry \cite{hsz,Franson89,Franson91} the spatial state of the two-particle system
is a coherent superposition of two distinct pairs of correlated paths for the two subsystems. In the same
spirit as Pancharatnam we use a two-particle interferometric intensity, namely the coincidence intensity
in a Franson interferometer \cite{Franson89,Franson91,Hessmo00}, to define an "in-phase" condition and the
corresponding parallel transport. An arbitrary unitary operation is performed in one of the arms of the
interferometer on the first subsystem, thus making the outputs of the two possible pairs of paths different.
Subsequently, another unitary operation is performed in the other arm on the second subsystem, and is
chosen such as to achieve maximal coincidence intensity. This second unitary is considered to be the
"phase" degree of freedom of the system and at maximal intensity the two outputs are considered to be
"in-phase", or "parallel". Thus we consider the orbit space formed by the state space of the system
modulo this unitary degree of freedom on the second subsystem to be the space in which the system is
parallel transported.

In the special case of pure two-qubit states, and evolutions generated by local $SU(2)$ operations,
the state space naturally fibrates through the second Hopf fibration and the orbit space of the two
qubit-states can be mapped to the state space of a quaternionic qubit \cite{mosseri01}. The coincidence
intensity of the Franson interferometer, in this case, corresponds to the quaternionic quantum mechanics
analogue of the Mach-Zehnder intensity and the associated parallel transport condition corresponds to
the L\'{e}vay connection \cite{levay04} restricted to local $SU(2)$ evolutions.

The outline of the paper is as follows. It turns out that the Stokes tensor formalism \cite{jaegger}
is convenient for our analysis. Therefore we briefly review this representation in section \ref{filt}.
Section \ref{punt} contains a description of the Franson interferometric setup and we define the
parallelity condition in this setting. In section $\ref{yuopi}$, we describe the parallel transport
procedure, discuss the properties of the related holonomy group and introduce the corresponding
connection form. Finally, in section \ref{gnutt} we consider the parallel transport scheme in the
special case of pure two-qubit states, and evolutions generated by $SU(2)$ operations and its
relation to the quaternionic representation of pure two-qubit states and the corresponding L\'{e}vay
connection. The paper ends with the conclusions.

\section{Stokes tensor formalism}

In the Stokes\label{filt} tensor formalism \cite{jaegger}, single particle quantum states are
represented as real vectors and $N$-partite states are represented as real $N$-tensors. A multi-partite
system consisting of parts $A,B,\dots{Z}$, where part $K$ has dimension $D_K$, is represented as a $D_{A}^2\times{D_{B}^2}\times\dots\times{D_{Z}^2}$-dimensional tensor $S_{j_{A}j_{B}\dots{j}_{Z}}$.
This tensor is related to the density matrix representation $\hat{\rho}$ of the same state as
\begin{eqnarray}
\hat{\rho}=\!\!\sum_{j_{A}=0}^{D_{A}^2-1}\!\sum_{j_{B}=0}^{D_{B}^2-1} \! \! \dots \! \!
\sum_{j_{Z}=0}^{D_{Z}^2-1} \! \! \left[\prod_{K=A}^{Z}\frac{1}{\delta_{0j_{K}}\!(D_{K} \! -2) \! + \!2}
\right] \! \! S_{j_{A} j_{B} \ldots j_{Z}} \hat{\chi}^{A}_{j_{A}} \! \otimes \! \hat{\chi}^{B}_{j_{B}} \! \otimes
\! \dots \! \otimes\!\hat{\chi}^{Z}_{j_{Z}},
\nonumber\\
\end{eqnarray}
where $\hat{\chi}^{K}_{0}\equiv{\hat{1}^{K}}$ and $\hat{\chi}^{K}_{j_{K}}$ $(j_{K}=1,2,...,D_{K}^2-1)$ are
the $D_{K}^2-1$ traceless generators of $U(D_{K})$ operations on system $K$. In the following, we shall
in most cases use the simplifying notation $\hat{\chi}_{j}^{K}$ for the generators on subsystem $K$.
The Hermitean, traceless and linearly independent generators $\{\chi_{j}^{K}\}_{j=1}^{D^2_{K}-1}$ of
$U(D_{K})$ satisfy orthogonality $\Tr(\hat{\chi}_{j}^{K}\hat{\chi}_{k}^{K})=2\delta_{jk}$, and
$[\hat{\chi}_{k}^{K},\hat{\chi}_{l}^{K}] = 2i \sum_{m=1}^{D^2_{K}-1}f_{klm}\hat{\chi}^{K}_{m}$
as well as $\{\hat{\chi}^{K}_{k},\hat{\chi}^{K}_{l}\}=\frac{4}{D_{K}}\delta_{kl}\hat{1}^{K} +
2 \sum_{n=1}^{D^2_{K}-1}d_{kln}\hat{\chi}^{K}_{n}$, where $d_{jkl}$ and
$f_{jkl}$ are the symmetric and antisymmetric structure constants, respectively, of
$U(D_{K})$. The factors $\left[ \delta_{0j_K} (D_K - 2) +2
\right]^{-1}$ are inserted so that $S_{j_Aj_B\ldots j_Z} = \Tr \left( \hat{\rho} \hat{\chi}^{A}_{j_{A}}
\! \otimes \! \hat{\chi}^{B}_{j_{B}} \! \otimes \! \dots \! \otimes\!\hat{\chi}^{Z}_{j_{Z}} \right)$.

We shall represent unitary operators $U(D_{K})$ on a form compatible with this formalism.
These are represented by complex vectors with elements $U_{j}=\frac{1}{[\delta_{0j}(D_{K}-2)+2]}\Tr(\hat{U}\hat{\chi}_{j}^{K})$ so that
\begin{eqnarray}
\hat{U}=\sum_{j=0}^{D_{K}^2-1}U_{j}\hat{\chi}^{K}_{j}.
\end{eqnarray}
Unitarity of the operators demand that $\{ U_j \}$ are complex numbers satisfying
\begin{eqnarray}
\mid \! {U_{0}} \! \mid^2 + \frac{2}{D_{K}} \sum_{j=1}^{D_{K}^2-1} \! \mid \! {U_{j}} \! \mid^2 = 1
\end{eqnarray} and
\begin{eqnarray}
\sum_{j,k=0}^{D_{K}^2-1} U_{j} U_{k}^{\ast} \left[d_{jkl} + if_{jkl} + (\delta_{j0}\delta_{kl} +
\delta_{k0}\delta_{jl}) \right] = 0
\end{eqnarray}
for each $1\leq{l}\leq{D_{K}^2-1}$.

In this paper we focus on bipartite systems $A+B$ and therefore it can be instructive to consider
some general properties of the Stokes 2-tensor. The zeroth row and zeroth column of the tensor, with
elements $S_{0j}$ and $S_{j0}$ respectively, are the Stokes 1-tensors corresponding to the reduced
states of subsystem $A$ and $B$, and these contain all the local information of the bipartite system.
The remaining part, formed by the elements $S_{ij},\phantom{u}i,j>0$, contains all the correlations
of the state and this subtensor, the correlation matrix, is denoted by $M_{ij}$. This matrix
has previously been used to study correlations and separability in quantum systems
\cite{horodecki1996,asa2002}.

A unitary transformation on subsystem $A$ transforms the reduced density matrix as
$\hat{U}\hat{\rho}^A\hat{U}^{\dagger}$. This corresponds to a transformation $RS$, on $S$ where $R_{jk}=\frac{1}{((D_{A}-2)\delta_{j0}+2)}\Tr(\hat{U}\hat{\chi}^A_{k}\hat{U}^{\dagger}\hat{\chi}^A_{j})$. The $R$ matrix is
an orthogonal matrix, as can be seen by observing that
\begin{eqnarray}
\label{ort}
\delta_{jk} \! \! & = & \! \! \frac{1}{[(D_{A}\!-\!2)\delta_{j0} \! + \!2]} \!\Tr(\hat{\chi}^A_{j}\hat{\chi}^A_{k})=\frac{1}{[(D_{A}\!-\!2)\delta_{j0}\!+\!2]} \!\Tr(\hat{U}\hat{\chi}^A_{j}\hat{U}^{\dagger}\hat{U}\hat{\chi}^A_{k}\hat{U}^{\dagger}) =
\nonumber\\
&=&\frac{1}{[(D_{A}\!-\!2)\delta_{k0}\!+\!2][(D_{A}\!-\!2)\delta_{j0}\!+\!2]} \! \! \sum_{l=0}^{D_{A}^2-1}\!(U\chi_{j}^AU^{\dagger})_{l}(U\chi_{k}^AU^{\dagger})_{l} = \! \!
\sum_{l=0}^{D_{A}^2-1}\!\!R^T_{jl}R_{lk}.
\nonumber\\
\end{eqnarray}
Similarly a unitary transformation on subsystem $B$ corresponds to an orthogonal matrix acting
on $S$ from the right. It should be noted that the $M$ matrix transforms under local unitary
operations on subsystem $A$ and $B$ through left and right action, respectively, by orthogonal
matrices.

As an example of how the $M$ matrix behaves we can consider a pure two-qubit state in the
Schmidt basis
\begin{eqnarray}
\mid \! \psi \rangle = a \! \mid \! {00} \rangle + b \! \mid \! {11} \rangle
\end{eqnarray}
where $a$ and $b$ are real non-negative numbers and $a^2+b^2=1$. Expressed in the Schmidt
basis, the correlation matrix with elements $M_{jk} = \Tr \left( \hat{\rho} \hat{\sigma}_j^A
\!\otimes\! \hat{\sigma}_k^B \right)$, $\sigma_j^A$ and $\sigma_k^B$, $j,k=1,2,3$, being
the standard Pauli operators on subsystem $A$ and $B$, respectively, reads
\begin{eqnarray}
M = \left(\begin{array}{ccc}
C & 0 & 0\\
0 & -C & 0\\
0 & 0 & 1
\end{array}\right),
\end{eqnarray}
where $C=2ab$ is the pure state concurrence \cite{woot}. Here it is clearly seen that $M$ is
rank one for product states. It must be emphasized that $M_{11}=-M_{22}=C$ because we choose
$a,b\in\mathbb{R}$. The elements of $M$ are not explicit functions of concurrence for arbitrary
complex coefficients $a,b$. However, since all pure two-qubit states with the same concurrence
can be related by local unitaries, corresponding to orthogonal transformations acting from the
right and the left on $M$, the absolute value of the determinant of $M$
\begin{eqnarray}
\mid\!\det{M}\!\mid = C^2
\end{eqnarray}
is invariant under local unitary transformations and measures concurrence. For maximally
entangled states we thus have that $\mid \! \! \det{M} \! \! \mid = 1$.

When we consider mixed states there can be correlations also in separable states. As an example
of how the $M$ matrix registers correlation for mixed states we can consider the Werner states
\cite{werner}
\begin{eqnarray}
\hat{\rho}_{W} = p\mid\!\psi\rangle\langle\psi\!\mid+\frac{(1-p)}{4} \hat{1}^{A}\!\otimes\hat{1}^{B},
\end{eqnarray}
where $\mid\!\!\psi\rangle$ is some maximally entangled state and $p\in[0,1]$. The absolute value
of the determinant is $\mid\!\det{{M}}\!\mid=p^3$, which can be compared to the square of the
concurrence $C=\max\left[\frac{(3p-1)}{2},0\right]$. Thus for $p\leq{\frac{1}{3}}$ the determinant
of $M$ is nonzero even though the state is separable. This underscores that the $M$ matrix for mixed
states is sensitive to correlations in general.

\section{Parallel transport condition}
\label{punt}
\subsection{Interferometric setup and parallelity}

A Franson interferometer \cite{Franson89,Franson91,Hessmo00},is a two-particle device composed of
two identical unbalanced two-path interferometers, so-called Franson loops, as shown in Fig. 1.
The two parts of the bipartite state are emitted, one into each Franson loop. A beam splitter
divides each path in two different paths of unequal length, which later converge at a second
beam splitter. The  difference in path length between the two arms in each Franson loop is
chosen to be the same, and such that the difference in transit time $\Delta{t}$ is greater
than the single particle coherence time. The transit time difference between the paths must
be smaller than the coherence time of the bipartite state to allow two-particle interference.
Of importance is that the emitter is such that is impossible to define a time of emission.
Detectors are placed after the convergence of the two paths in each Franson loop and coincidence
measurements are made.

\begin{figure}[h]
\includegraphics[width=12 cm]{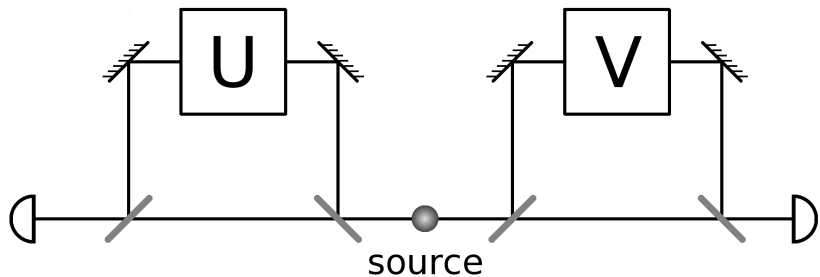}
\caption{\footnotesize{The Franson interferometer setup. Unitary operations on the
internal degrees of freedom the particle are performed in the long arms of each Franson
loop.}}
\end{figure}

Since we \label{ghgh}discard non-coincidental detections, corresponding to the bipartite
system traversing a long path in one of the Franson loops and a short path in the other,
it is necessary that the time resolution of the detectors is smaller than $\Delta{t}$.
Furthermore, due to the requirement that time of emission cannot be defined, and since no
measurements are made inside the Franson loops, it cannot be ascribed to a coincidence
detection event that the bipartite system traversed either the two short paths or the two
long ones. Hence the system is in a coherent superposition of having traversed the two
long arms, and having traversed the two short arms. In the long paths of each sub-interferometer
we place devices that perform unitary operations on the internal state of the bipartite state.
After the point of convergence of the two paths, the effective unnormalized internal state
is under the above requirements therefore

\begin{eqnarray}
\hat{\rho}=\frac{1}{4}(\hat{1}+\hat{U}\!\otimes\!\hat{V})\hat{\rho}_{0}
(\hat{1}+\hat{U}^{\dagger}\!\otimes\!\hat{V}^{\dagger}),
\end{eqnarray}
where $\hat{\rho}_{0}$ is the initial internal state. Given $\hat{U}=\sum_{j=0}^{D_{A}^2-1}
U_j\hat{\chi}_{j}^A$, and $\hat{V}=\sum_{j=0}^{D_{B}^2-1}V_j\hat{\chi}_{j}^B$ where
$\hat{\chi}_{j}^{A}$ and $\hat{\chi}_{j}^{B}$ are the generators of $U(D_{A})$ and $U(D_{B})$,
respectively, the coincidence detection intensity $I^{AB}$ is

\begin{eqnarray}
I^{AB}={\frac{1}{2}+\frac{1}{2}\re\Tr(\hat{U}\otimes\hat{V}\hat{\rho}_{0})}.
\end{eqnarray}
The coincidence detection intensity $I^{AB}$, henceforth referred to simply as "intensity",
is the ratio between the measured intensity for a given $\hat{U}$ and $\hat{V}$, and the
intensity measured if $\hat{U}=\hat{V}=\hat{1}$. The expression for $I^{AB}$ in the Stokes
tensor formalism is
\begin{eqnarray}
I^{AB}=\frac{1}{2}+\frac{1}{2} \! \sum_{k=0}^{D_{A}^2-1} \sum_{j=0}^{D_{B}^2-1}
\re(V_{j}U_{k})S_{kj},\label{asna}
\end{eqnarray}
where the second term is the interference term.

We now define the parallelity condition in the Franson setup for a bipartite system consisting
of two qudits of dimension $D_{A}$ and $D_{B}$. We ask, given that a \label{pi}specific unitary
operation $\hat{U}\in{U(D_{A})}$ has been chosen in the first Franson loop, what unitary
operation $\hat{V}\in{U(D_{B})}$ should be chosen in the second Franson loop in order to
maximize the coincidence intensity? We take maximal coincidence intensity as the definition
of parallelity between the output of the two short paths and the output of the two long paths.
This maximization procedure is the analogue of the procedure used to define Pancharatnam
parallelity in the context of a Mach-Zehnder interferometer \cite{wash1,wash2}, but here
the Franson coincidence intensity has taken the role of the Mach-Zehnder intensity and
$\hat{V}$ has taken the role of the $U(1)$ phase factor. It should be noted that if $S$
does not have full rank, then there exist $\hat{U}$ such that $\sum_{k=0}^{D_{A}^2-1} U_{k}
S_{kj}=0$ for all $j$. In this case the interference term is identically zero for all
$\hat{V}$, but if $\sum_{k=0}^{D_{A}^2-1}U_{k}S_{kj}\neq0$ there will always be a $\hat{V}$
corresponding to maximal intensity.

To find a formal expression for the operator $\hat{V}$ that maximizes the intensity, we
seek to maximize $I^{AB}$ in equation (\ref{asna}) with respect to the coefficients $V_{k}$,
using Lagrange's method. To enforce unitarity of $\hat{V}$ we introduce the constraints $\mid\!\!{V}_{0}\!\!\mid^2+\frac{2}{D_{B}}\sum_{j=1}^{D_{B}^2-1}\mid\!\!{V}_{j}\!\!\mid^2=1$
and $\sum_{j,k=0}^{D_{B}^2-1}V_{j}V_{k}^{\ast} \left[d_{jkl}+if_{jkl}+(\delta_{j0}\delta_{kl} +
\delta_{k0}\delta_{jl})\right]=0$ for each $1\leq{l}\leq{D_{B}^2-1}$. We thus construct the 
auxiliary function $f(\{V_{j}\},\{V_{j}^*\})$, that is to be extremized, as
\begin{eqnarray}
f\!\left(\{V_{j}\},\{V_{j}^*\}\right) = \sum_{k=0}^{D_{A}^2-1} \sum_{j=0}^{D_{B}^2-1}
(V_{j}U_{k}+V_{j}^*U_{k}^*)S_{jk}
\nonumber\\
-\lambda\!\!\left(V_{0}V_{0}^*+\frac{2}{D_{B}}\sum_{j=1}^{D_{B}^2-1}V_{j}V_{j}^*-1\right)
\nonumber\\
-\sum_{l=1}^{D_{B}^2-1}\mu_{l}\sum_{j,k=0}^{D_{B}^2-1}V_{j}V_{k}^{\ast} \left[d_{jkl} + if_{jkl} +
(\delta_{j0}\delta_{kl}+\delta_{k0}\delta_{jl}) \right].
\nonumber \\
\end{eqnarray}
Using Lagrange's method we seek the points were the gradient of the auxiliary function
with respect to the variables $V_{j}$ and $V_{j}^*$ vanishes. The components $V_{k}$
defining these points satisfy the equations
\begin{eqnarray}
\label{tyfus}{\lambda}
\! \left[\frac{2}{D_{B}} \! - \! \! \left( 1 \! - \! \frac{2}{D_{B}}\right) \! \delta_{k0}\right] \! \! {V_{k}}+\!\!\sum_{l=1}^{D_{B}^2-1}\!\!\mu_{l}\!\!\left[\!(\delta_{kl}V_{0}+\delta_{k0}V_{l})+ \! \! \sum_{j=1}^{D_{B}^2-1}\!\!V_{j}(d_{jkl}+if_{jkl})\!\right]\!\!\!=
\nonumber \\ = - \! \sum_{j=0}^{D_{A}^2-1}\!U_{j}^*S_{jk},
\phantom{yuyuyuyuy}0\leq{k}\leq{D_{B}^2-1}.
\end{eqnarray}
If the \label{pentagon} coefficients $U_{j}$ are ordered as a $D_{A}^2$-dimensional vector
$\bar{u}$ and likewise the coefficients $V_{k}$ are ordered as a $D_{B}^2$-dimensional vector
$\bar{v}$, the above equations can be reexpressed as a matrix equation
\begin{eqnarray}
B\bar{v} = -S^{T} \bar{u}^{\ast} ,
\end{eqnarray}
where $B$ is a $\lambda$ and $\mu_{l}$ dependent Hermitean matrix, given by
\begin{eqnarray}
B_{kl} = \lambda \! \! \left[ \frac{2}{D_{B}} \delta_{kl} - \! \left( 1-\frac{2}{D_{B}} \right) \!
\delta_{l0} \delta_{k0}\right] + \! \! \sum_{j=1}^{D_{B}^2-1} \! \mu_{j} ( \delta_{jk} \delta_{l0} +
\delta_{jl} \delta_{0k}+d_{jlk}-if_{jkl}).
\nonumber\\
\end{eqnarray}
Provided $B(\lambda,\bar{\mu})$ is invertible, the formal solution for $\hat{V}$ can be given as

\begin{eqnarray}
\hat{V} = \sum_{l,k,j=0}^{D^2_{B}-1} B^{-1}_{lk} (\lambda,\bar{\mu}) S^{T}_{kj}U_{j}^{\ast}
\hat{\chi}^{B}_{l},
\label{uuw}
\end{eqnarray}
where $\bar{\mu}\equiv\!\{\mu_{j}\}$. The explicit form of the Lagrange parameters, and thus
$B^{-1}$, is found by solving for the unitarity constraints on $\hat{V}$. The solutions of
these constraint equations give us the critical points of the intensity as a function of
$\hat{U}$. We can see from the constraints that for each solution $\lambda,\bar{\mu}$ there
is a solution $-\lambda,-\bar{\mu}$ and if one of them corresponds to a local maximum, the
other corresponds to a local minimum. There will be a unique solution to the maximization
problem if and only if there is a unique global maximum of the intensity as a function of
$\hat{U}$, corresponding to a combination of parameters $\lambda,\bar{\mu}$ satisfying the
constraints.

In the general case, finding this solution as a function of $\hat{U}$ appears to be a non-trivial problem. For product
states, however, the unitary $\hat{V}$ that maximizes the intensity is easily found and is
always a Abelian $U(1)$ phase factor. This can be seen by observing that for product states
$\rho^{A}\!\otimes\!\rho^{B}$ the coincidence\label{uuty} intensity is
\begin{eqnarray}
I^{AB}={\frac{1}{2}+\frac{1}{2}\re\Tr(\hat{U}\hat{\rho}^{A})\Tr(\hat{V}\hat{\rho}^{B})}.
\end{eqnarray}
and therefore the $\hat{V}$ that maximizes this expression is found to be
$\hat{V}=e^{-i\arg({\Tr(\hat{U}\hat{\rho}^{A})})}\hat{1}$. \label{pwing}
We \label{tube} may note that for the trivial case where $\hat{U}=e^{i\phi}\hat{1}$,
we see that the intensity is maximal if and only if $\hat{V}=e^{-i\phi}\hat{1}$, regardless
of the state of the bipartite system.

\subsection{Example I: Qudit-qubit}
In section \ref{pentagon}, the solution $\hat{V}\in{U(D_{B})}$ to the maximization problem
is not given on a closed form and it is not apparent how to find it. However, for the case
where the second subsystem is a qubit, and therefore $\hat{V}\in{U(2)}$, the $B$ matrix is $B_{kl}={\lambda}\delta_{lk}+\sum_{j=1}^{3}\mu_{j}\left[(\delta_{jk}\delta_{l0}+\delta_{jl}
\delta_{k0})-if_{jkl}\right]$. It can be seen that $B$ now separates as $B=\lambda\hat{1}+H$,
where $H$ is Hermitean and $H^2\propto\hat{1}$, and therefore $B^{-1}=\frac{1}{\sqrt{\det{B}}}
(-\lambda\hat{1}+H)$. Explicitly
\begin{eqnarray}
B^{-1} & = & \frac{1}{\sqrt{\det{B}}}\left(\begin{array}{cccc}
-{\lambda} & \mu_{1} & \mu_{2} & \mu_{3} \\
\mu_{1} & -\lambda & -i\mu_{3} & i\mu_{2} \\
\mu_{2} & i\mu_{3} & -{\lambda} & -i\mu_{1} \\
\mu_{3} & -i\mu_{2} & i\mu_{1} & -{\lambda}
\end{array}\right),
\nonumber \\
\det{B} & = & \left(-\lambda^2+\bar{\mu}\cdot\bar{\mu}\right)^2.
\label{hjko}
\end{eqnarray}
While it is still not obvious how to find the Lagrange parameters in general we may solve it
in some special cases. To illustrate this, let us consider a $\hat{U}$ that has the form
$\hat{U}=U_{1}\hat{\sigma}_{1}+U_{2}\hat{\sigma}_{2}$, where $\hat{\sigma}_{1}$ and
$\hat{\sigma}_{2}$ are the standard Pauli operators, and a pure state
\begin{eqnarray}
\mid\!\psi\rangle=a\!\mid\!{00}\rangle+b\!\mid\!{11}\rangle,
\end{eqnarray}
where $a,b\geq{0}$. As a consequence of our choice of basis  $S_{10}=S_{20}=0$ and the
concurrence $C=2ab$. This together with our special choice of $\hat{U}$, implies that the
intensity is
\begin{eqnarray}
I^{AB} = \frac{1}{2} + \frac{C}{2} \re(V_{1}U_{1}-V_{2}U_{2}).
\end{eqnarray}
Using equations (\ref{uuw}) and (\ref{hjko}), and the constraints, we find that
$\bar{\mu}=0,\lambda=\pm{C}$, where the positive sign corresponds to the global maximum while
the negative sign corresponds to the global minimum. The  unitary operator $\hat{V}$ corresponding
to the maximum is
\begin{eqnarray}
\hat{V}=U_{1}^*\hat{\sigma}_{1}-U_{2}^*\hat{\sigma}_{2},
\end{eqnarray}
and the maximal intensity is
\begin{eqnarray}
I^{AB}_{max} = \frac{1}{2}(1+{C}).
\end{eqnarray}

\subsection{Example II: Restriction to $SU(D)$}
A variation of the maximization procedure is to restrict the set from which $\hat{U}$ and $\hat{V}$
can be chosen. One natural restriction would be to consider only $SU(D_{A})$ and $SU(D_{B})$ operations
in the Franson loops. When this restriction is made the qualitative properties of the parallel
transport may change. It is for example no longer obvious that the unitaries $\hat{V}$ associated
to a product state, will be a commuting set in the general case. The restriction where the second
subsystem is a qubit and therefore $\hat{V}\in{SU(2)}$, however, leads to a significant simplification
of the maximization problem. Here, we solve this problem and in particular show that product states
are indeed associated with commuting sets of unitaries.

Since $SU(2)$ can be parametrized by four real numbers subject to only one constraint, the solution
of the maximization problem can be found easily for arbitrary states and arbitrary $\hat{U}\in{SU(D_{A})}$.
We choose the parametrization of $\hat{V}$ such that $V_{0}$ is real and $iV_{1},iV_{2},iV_{3}$ are
purely imaginary. The intensity in this parametrization is
\begin{eqnarray}
\label{label}
I^{AB}=\frac{1}{2}+\frac{1}{4}\left[\sum_{j=0}^{D_{A}^2-1}V_{0}(U_{j}+U_{j}^*)S_{j0} +
i\sum_{j=0}^{D_{A}^2-1}\sum_{k=1}^{3}V_{k}(U_{j}-U_{j}^*)S_{jk}\right]
\end{eqnarray}
and $\hat{V}$ is found to be
\begin{eqnarray}
\label{tgb}
\hat{V}=\frac{1}{2\lambda}\left[\sum_{j=0}^{D_{A}^2-1}(U_{j}+U_{j}^*)S_{j0}\hat{1}^{B} +i\sum_{j=0}^{D_{A}^2-1}\sum_{k=1}^{3}(U_{j}-U_{j}^*)S_{jk}\hat{\sigma}^{B}_{k}\right],
\end{eqnarray}
where $\hat{\sigma}^{B}_{m}$ are the Pauli operators. The remaining Lagrange parameter $\lambda$
is found from the unitarity condition $\sum_{j=0}^{3}\mid\!{V_{j}}\!\mid^2=1$ and is
\begin{eqnarray}
\lambda=\pm\frac{1}{2}\sqrt{\left[\sum_{j=0}^{D_{A}^2-1}(U_{j}+U_{j}^*)S_{j0}\right]^2 + \sum_{k=1}^{3}\left[\sum_{j=0}^{D_{A}^2-1}(U_{j}-U_{j}^*)S_{jk}\right]^2}.
\end{eqnarray}
The sign of $\lambda$ must be chosen positive since the trivial case $\hat{U}=\hat{1}$ implies
$\hat{V}=\hat{1}$.

When the bipartite state is a product state $\hat{\rho}=\hat{\rho}^{A}\!\otimes\!\hat{\rho}^{B}$
we find that $\hat{V} = \frac{1}{2\lambda}[\sum_{j=0}^{D_{A}^2-1}(U_{j}+U_{j}^*)S_{j0}\hat{1}^{B} +2i\sum_{j=0}^{D_{A}^2-1}(U_{j}-U_{j}^*)S_{j0}(\hat{\rho}^{B}-\hat{1}^{B})]$. Hence for product states
the unitary $\hat{V}$ that maximizes intensity commutes with $\hat{\rho}^{B}$ for any $\hat{U}$.
Therefore the set of unitaries $\hat{V}$ associated to a product state commute with the density
operator and with each other. The signature of a product state is thus that the corresponding set
of unitaries will be commuting when the unitary operators on the second subsystem are restricted
to $SU(2)$.

\subsection{Example III: SO(D) and two-redit states}
Another variation of of our procedure is to maximize the intensity for $\hat{V}\in{SO(D_{B})}$
given $\hat{U}\in{SO(D_{A})}$ in the other Franson loop.
Since $SO(D)$ is the
endomorphism group of the $D$-dimensional redit state space we may consider this restriction of
the maximization procedure when the state space is restricted to a two-redit subspace. For two
redits the state $\hat{\rho}$ naturally decomposes as
\begin{eqnarray}
\hat{\rho} = \frac{1}{[\delta_{k0}(\!D_{A}\!-\!2)\!+\!2][\delta_{l0}(\!D_{B}\!-\!2)\!+\!2]}
\sum_{k=0}^{D_{A}^2-1}\!\sum_{l=0}^{D_{B}^2-1} \!\left(S^{sym}_{kl}+S^{anti-sym}_{kl}\right)\hat{\chi}_{k}^A \! \otimes \! \hat{\chi}^B_{l},
\nonumber\\
\end{eqnarray}
where $S^{sym}_{kl}$ is nonzero only when $\hat{\chi}_{k}^{A}$ and $\hat{\chi}_{l}^{B}$ are both symmetric, and $S^{anti-sym}_{kl}$ is nonzero only when $\hat{\chi}_{k}^{A}$ and $\hat{\chi}_{l}^{B}$ are both antisymmetric.
The generators of the subgroup $SO(D)\subset{SU(D)}$ are the antisymmetric
generators of $SU(D)$. As a special case we can consider a general mixed two-rebit state,
where the only pair of antisymmetric generators spanning the state space is $\hat{\sigma}_{2}^A \otimes
\hat{\sigma}_{2}^B$. Since $SO(2)$ operators can be expanded in a basis consisting of only $\hat{1}$ and $\hat{\sigma}_{2}$, the intensity is
\begin{eqnarray}
I^{AB}=\frac{1}{2}+\frac{1}{2}U_0V_0-\frac{1}{2}V_{2}U_{2}M_{22},
\end{eqnarray}
where $U_{0}, V_{0}, U_{2}, V_{2}\in\mathbb{R}$ and $M_{22}=C_{R}=\Tr(\hat{\sigma}_{2}^A \otimes
\hat{\sigma}_{2}^B\hat{\rho})$ is the rebit concurrence \cite{cavesfuchsrungta}. The $SO(2)$ operator
that maximizes the intensity is
\begin{eqnarray}
\hat{V}=\frac{1}{\lambda}\left(U_{0}\hat{1}^{B}-iC_{R}U_{2}\hat{\sigma}^{B}_{2}\right),
\end{eqnarray}
where $\lambda=\sqrt{\mid\!{U_{0}}\!\mid^2+C_{R}^2\!\mid\!{U_{2}}\!\mid^2}$. We note that for product two-rebit states $\hat{V}$ can only be $\hat{1}$ or $-\hat{1}$.

\section{Correlation induced non-Abelian quantum holonomy}
\label{yuopi}
In this section we use the parallelity condition introduced in section \ref{uuty} to define a procedure
for parallel transport of a bipartite quantum state. We consider the infinitesimal limit to find a
connection form corresponding to this parallel transort. By definition the output of the long arms is
parallel with the output of the short arms when $\hat{V}$ is chosen such as to maximize the coincidence
intensity. Now we choose to view the output of the long arms as the parallel transported version of the
output of the short arms. By using that output state as input for another Franson setup, where in a
similar way a new output state is created, we can parallel transport the state through an arbitrary
number of steps.

To see how this works we let $\hat{\rho}^{(0)}$ be the input state of the interferometer. The output
state of the long arms is $\hat{\rho}^{(1)} = {\hat{U}^{(1)} \otimes{V}^{(1)}} \hat{\rho}^{(0)}
\hat{U}^{(1)\dagger} \otimes{V}^{(1)\dagger}$, where $\hat{U}^{(1)}\in{U(D_{A})}$ and $\hat{V}^{(1)}
\in U(D_{B})$ are unitary operators that has been applied such as to implement parallelity.
In the second step, we use $\hat{\rho}^{(1)}$ as the input in a new Franson interferometer, where a
new unitary $\hat{U}^{(2)}$ is chosen and a new $\hat{V}^{(2)}$ is found to create an output of the
long arms $\hat{\rho}^{(2)}$ that is parallel to $\hat{\rho}^{(1)}$.

The parallel transport is performed by iterating the intensity maximizing procedure in this way as
illustrated in Fig. 2. In the $n$th step, a $\hat{U}^{(n)}\in{U(D_{A})}$ is chosen, and thereafter
a $\hat{V}^{(n)}\in{U(D_{B})}$ is found that maximizes the intensity. After $\hat{V}^{(n)}$ has
been found the input state for the next step is taken to be $\hat{\rho}^{(n)} = (\hat{U}^{(n)} \!
\otimes \! \hat{V}^{(n)})\hat{\rho}^{(n-1)}( \hat{U}^{(n)\dagger} \! \otimes \! \hat{V}^{(n)\dagger})$.

\begin{figure}[h]
\includegraphics[width = 12 cm]{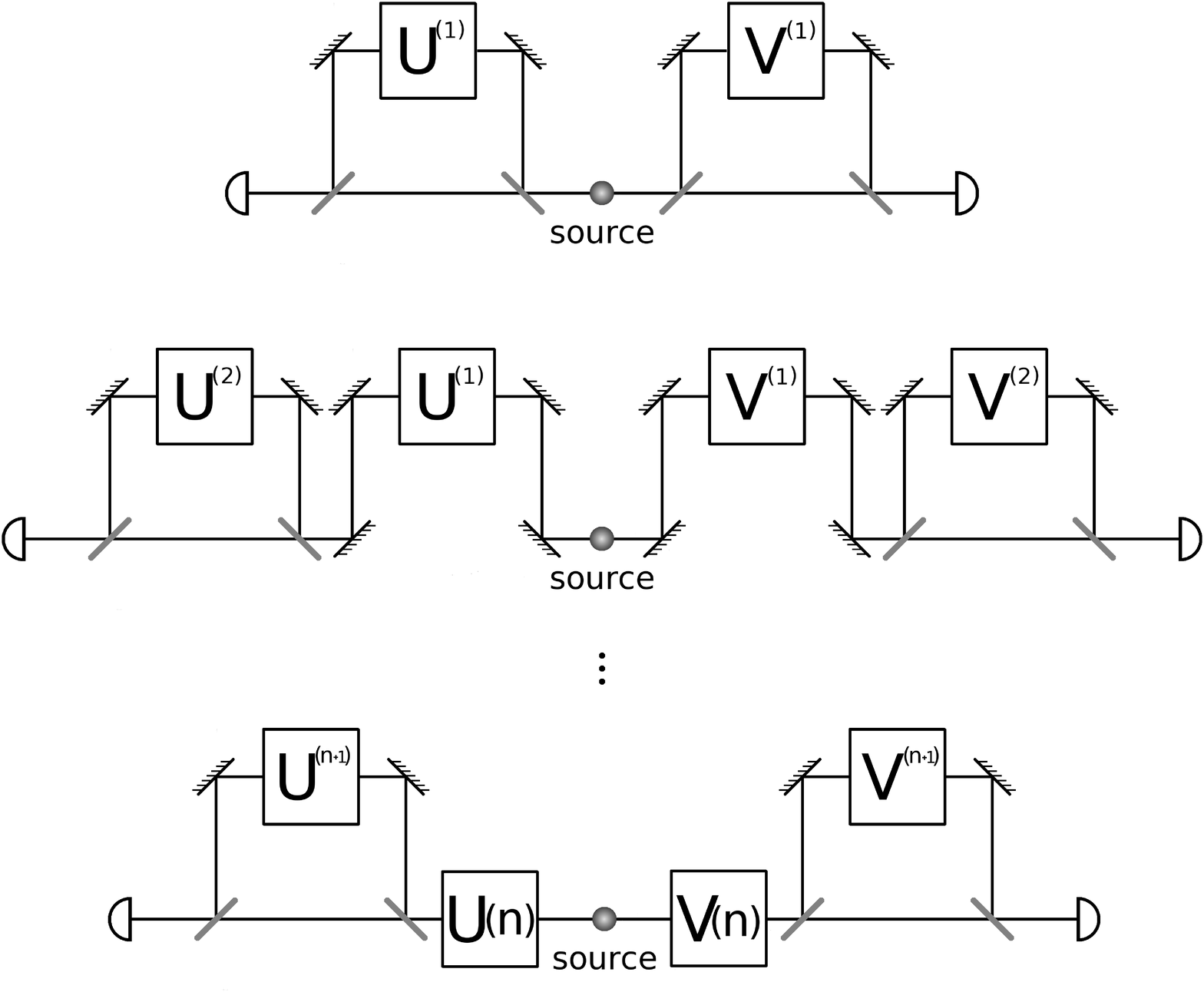}
\caption{\footnotesize{Iterative procedure in first, second and $(n+1)$th step.}}
\end{figure}
The coincidence intensity in the $(n+1)$th step is now

\begin{eqnarray}
I^{(n+1)}=\frac{1}{2}+\frac{1}{2}\re\Tr\left[(\hat{U}^{(n+1)}\!\otimes\! \hat{V}^{(n+1)})({\hat{U}(n)\!\otimes\!\hat{V}(n)})\hat{\rho}^{(0)}({\hat{U}(n)^{\dagger}
\!\otimes\!{\hat{V}(n)^{\dagger}}})\right],
\nonumber\\
\end{eqnarray}
where the cumulated unitary operations that are applied to the original input state
$\hat{\rho}^{(0)}$ at the beginning of the $(n+1)$th step are $\hat{U}(n)\equiv{\hat{U}^{(n)}
\hat{U}^{(n-1)}}\dots{\hat{U}^{(1)}}$ and $\hat{V}(n)\equiv{\hat{V}^{(n)}\hat{V}^{(n-1)}
\dots{\hat{V}^{(1)}}}$.

From this we can define a holonomy group $Hol_{S}$ based on a particular state $\hat{\rho}_{0}$
with corresponding Stokes tensor $S$ as the set of unitary operators $\hat{V}(n)\equiv{\hat{V}^{(n)}
\hat{V}^{(n-1)}\dots{\hat{V}^{(1)}}}\in{U(D_{B})}$ that can result from the above parallel transport
prescription given all sequences of $\hat{U}^{(k)}\in{U(D_{A})}$ such that $\hat{U}(n) \equiv \hat{U}^{(n)} \hat{U}^{(n-1)}\dots{\hat{U}^{(1)}}=\hat{1}$ for any $n$.  From the discussion on product states at the
end of section \ref{pwing} follows that the holonomy group for product states is always Abelian, and
only correlated states can induce a non-Abelian holonomy group.

For any set of unitaries $\{\hat{U}^{(n)}\in{U}(D_{A})\}$ given by
\begin{eqnarray}
\hat{U}^{(n)}=U^{(n)}_{0}\hat{1}^{A}+\sum_{k=1}^{D_{A}^2-1}U^{(n)}_{k}\hat{\chi}^A_k,
\end{eqnarray}
we find $\hat{V}^{(n)}\in{U(D_{B})}$ that maximizes the intensity as
\begin{eqnarray}
 \label{plott}
\hat{V}^{(n)} = \sum_{j,k=0}^{D_{B}^2-1}\sum_{l=0}^{D_{A}^2-1}B^{-1}_{jk}(\lambda^{(n)},\bar{\mu}^{(n)})
S_{kl}^{T}U^{(n)*}_{l} \hat{\chi}^B_{j},
\nonumber\\
\end{eqnarray}
where $S_{lk}=\Tr\left[{(\hat{\chi}^A_{l}\!\otimes\!\hat{\chi}^B_{k})({\hat{U}(n-1)\!\otimes \!\hat{V}(n-1)})\hat{\rho}^{0}(\hat{U}(n-1)^{\dagger}\!\otimes\!\hat{V}(n-1)^{\dagger}})\right]$.

To define a connection we need to consider the limit when $\hat{U}$ and $\hat{V}$ are infinitesimally
close to unity. To find this limit we revisit the maximization problem with a different parametrization $\hat{U}=e^{i\sum_{j=0}^{D_{A}^2-1}\theta_{j}\hat{\chi}_{j}^{A}}$ and $\hat{V}=e^{i\sum_{j=0}^{D_{B}^2-1}
\phi_{j}\hat{\chi}^{B}_{j}}$. The intensity is then
\begin{eqnarray}
I^{AB}={\frac{1}{2}+\frac{1}{2}\re\Tr(e^{i\sum_{j=0}^{D_{A}^2-1}\!\!\theta_{j}\hat{\chi}_{j}^{A}}\otimes e^{i\sum_{j=0}^{D_{B}^2-1}\!\!\phi_{j}\hat{\chi}^{B}_{j}}\hat{\rho}_{0})},\end{eqnarray}
where $\theta_{j}$ and $\phi_{j}$ are real numbers. In this representation unitarity is explicit and no
constraints are necessary. Differentiating $I^{AB}$ with respect to the parameters $\phi_{j}$ and setting
each derivative to zero, we find
\begin{eqnarray}
\re\Tr(e^{i\sum_{k=0}^{D_{A}^2-1}\!\!\theta_{k}\hat{\chi}^{A}_{k}}\otimes{i}\hat{\chi}^{B}_{j}
{e}^{i\sum_{l=0}^{D_{B}^2-1}\!\! \phi_{l}\hat{\chi}^{B}_{l}}\hat{\rho}_{0})=0,\phantom{uuuuu}
0 \leq j \leq D_{B}^2-1.
\nonumber\\
\end{eqnarray}
Since we are only interested in finding the connection form we expand these equations to linear order in
$\theta_{j}$ and $\phi_{j}$, to obtain
\begin{eqnarray}
\re\Tr\!\left[\!\left(\hat{1}^{A}\!+i\sum_{k=0}^{D_{A}^2-1}\!\theta_{k}\hat{\chi}^{A}_{k}
\right) \! \! \otimes\!{i\hat{\chi}^{B}_{j}}\! \! \left( \hat{1}^{B} \!
+i \! \! \sum_{l=0}^{D_{B}^2-1} \! \phi_{l}\hat{\chi}^{B}_{l} \right) \!\hat{\rho}_{0}\right]\! =
\nonumber\\
=\!\left(\!-\left\{\!\sum_{k=0}^{D_{B}^2-1}\!\!\phi_{k}S_{0k}\delta_{j0}+\phi_{0}S_{0j}\!+\!\left[\frac{2}{D_{B}}+ \!\left(\!1-\frac{2}{D_{B}}\!\right)\!\!\delta_{j0}\right]\!\!\phi_{j}\!\right\}\!-\!\!\sum_{k,l=1}^{D_{B}^2-1} \! \!
\phi_{k}d_{jkl} S_{0l}  \right) \nonumber\\-\sum_{k=0}^{D_{A}^2-1}\theta_{k}S_{kj}=0,
\phantom{uuuuuuuuuuu} 
0 \leq j \leq D_{B}^2-1.
\end{eqnarray}
In the infinitesimal limit we introduce the notation $d\hat{U}\hat{U}^{\dagger}\equiv{i}(\sum_{j=0}^{D_{A}^2-1}d
\theta_{j}\hat{\chi}^{A}_{j})$ and likewise $d\hat{V}\hat{V}^{\dagger}\equiv{i}(\sum_{j=0}^{D_{A}^2-1}d\phi_{j}
\hat{\chi}^{A}_{j})$. Although performing infinitesimal unitary operations is clearly an idealization we may still
consider this limit where the sequences of unitaries $\{\hat{\Delta}_{U}^{(t)}\equiv{\hat{1}^{A}+d\hat{U}
\hat{U}^{\dagger}(t)}\}$ and $\{\hat{\Delta}_{V}^{(t)}\equiv{\hat{1}^{B}+d\hat{V}\hat{V}^{\dagger}(t)}\}$
in the parallel transport are indexed by a continuous variable $t$.

The relation between $\{\hat{\Delta}_{U}^{(t)}\}$ and $\{\hat{\Delta}_{V}^{(t)}\}$, for each $t$, is given by
\begin{eqnarray}
\label{khuy}
\sum_{k=0}^{D_{B}^2-1}B_{jk}({dVV}^{\dagger})_{k}(t) =
-\sum_{k=0}^{D_{A}^2-1}({dUU}^{\dagger})_{k}(t)S_{kj},
\end{eqnarray}
where $B_{jk}$ are the elements of the symmetric matrix $B$, given as
\begin{eqnarray}
B_{jk}&=&\re\Tr(\hat{\chi}^{B}_{j}\hat{\chi}^{B}_{k}\hat{\rho}^{B}) =
\nonumber\\
&=&\left[\frac{2}{D_{B}}\delta_{kj}+\left(1-\frac{2}{D_{B}}\right)\delta_{j0}\delta_{k0}\right]
+ \sum_{l=1}^{D_{B}^2-1}S_{0l} \left[(\delta_{j0}\delta_{lk}+\delta_{k0}\delta_{lj})+d_{jkl}\right].
\nonumber\\
\end{eqnarray}
Therefore, provided $B$ is invertible, we find
\begin{eqnarray}
\hat{\Delta}_{V}\label{hot} = \hat{1}^{B}-\sum_{k=0}^{D_{A}^2-1}\sum_{l,m=0}^{D_{B}^2-1}
(dUU^{\dagger})_{k}S_{kl}{B}^{-1}_{ml}\hat{\chi}^{B}_{m}.
\end{eqnarray}
We can now identify
\begin{eqnarray}
\label{gbnm}
\hat{A}(t)\equiv-\sum_{k=0}^{D_{A}^2-1}\sum_{l,m=0}^{D_{B}^2-1}(dUU^{\dagger})_{k}S_{kl}{B}^{-1}_{ml}\hat{\chi}^{B}_{m}
\end{eqnarray}
as the operator-valued anti-Hermitean connection one-form. Note that $\hat{A}$ is a linear function of the Stokes
matrix $S$. If we decompose the density operator as $\hat{\rho}=\sum_{\mu}p_{\mu}\mid\!\psi_{\mu}\rangle\langle\psi_{\mu}\!\mid$
we can express the connection form as

\begin{eqnarray}
\hat{A}(t)=i\sum_{\mu}p_{\mu}\!\!\!\sum_{j,l=0}^{D_{B}^2-1}\!\!\re\!\Tr(\hat{1}^{A}\!\otimes\!i\hat{\chi}^{B}_{j}\mid\!d{\psi}_{\mu}(t) \rangle\langle\psi_{\mu}(t)\!\mid)B^{-1}_{jl}\hat{\chi}^{B}_{l}.
\end{eqnarray}
Under a change of gauge $\mid\!{\psi}_{k}\rangle\to{\hat{1}^A\!\otimes\!\hat{G}}\!\mid\!{\psi}_{k}\rangle$, corresponding to a unitary
transformation on the second subsystem, the connection transforms as
\begin{eqnarray}
\hat{A}\!\to\!\!\hat{A}^{'}\!\!=\!i\!\sum_{\mu}p_{\mu}\!\!\!\!\!\!\sum_{j,l,m,n=0}^{D_{B}^2-1}\!\!\!\!\!\!\re\!\Tr[\hat{1}^{A}\!\!\otimes\!i\hat{\chi}^{B}_{j}\!\!\mid\!\!d(\hat{1}^A\!\!\otimes\!\hat{G}{\psi}_{\mu})\rangle\langle\psi_{\mu}\hat{1}^A\!\!\otimes\!\hat{G}^{\dagger}\!\!\mid]R_{jl}B^{-1}_{lm}R^{T}_{mn}\hat{\chi}^{B}_{n}\!\!=
\nonumber\\
=i\!\sum_{\mu}p_{\mu}\!\!\!\!\!\!\sum_{j,l,m,n=0}^{D_{B}^2-1}\!\!\!\!\!\!\re\!\Tr[\hat{1}^{A}\!\otimes\!\hat{G}^{\dagger}i\hat{\chi}^{B}_{j}(\hat{G}\mid\!d{\psi}_{\mu}\rangle + d\hat{G}\mid\!{\psi}_{\mu}\rangle)\langle\psi_{\mu}\!\mid]R_{jl}B^{-1}_{lm}R^{T}_{mn}\hat{\chi}^{B}_{n} =
\nonumber\\
=i\!\sum_{\mu}p_{\mu}\!\!\!\!\!\!\sum_{h,j,l,m,n=0}^{D_{B}^2-1}\!\!\!\!\!\!\re\!\Tr[ \hat{1}^{A} \! \otimes \! i\hat{\chi}^{B}_{h}
\mid\!{d\psi}_{\mu} \rangle \langle\psi_{\mu} \! \mid]R^{T}_{hj}R_{jl}B^{-1}_{lm}R^{T}_{mn}\hat{\chi}^{B}_{n}
\nonumber\\
+i\!\sum_{\mu}p_{\mu}\!\!\!\!\!\!\sum_{h,j,l,m,n=0}^{D_{B}^2-1}\!\!\!\!\!\!\re\!\Tr[\hat{1}^{A}\!\otimes\!i\hat{\chi}^{B}_{h}\hat{G}^{\dagger}d\hat{G} \mid\!{\psi}_{\mu}\rangle\langle\psi_{\mu}\!\mid]R^{T}_{hj}R_{jl}B^{-1}_{lm}R^{T}_{mn}\hat{\chi}^{B}_{n} =
\nonumber\\ = \hat{G}d\hat{G}^{\dagger} + \hat{G}\hat{A}\hat{G}^{\dagger},
\end{eqnarray}
where $R_{jk}=\frac{1}{[\delta_{k0}(D_{B}-2)+2]}\Tr(\hat{G}\hat{\chi}_{k}^{B}\hat{G}^{\dagger}\hat{\chi}_{j}^{B})$, 
and we have used that $B_{jk}=\re\Tr(\hat{\chi}^{B}_{j}\hat{\chi}^{B}_{k}\hat{\rho}^{B})$ which transforms as 
$B\to{RBR^{T}}$. Thus $\hat{A}$ transforms as a proper gauge potential.

For a given path  $\gamma$ in $U(N_{A})$ given by $\hat{U}(t)$, the parallel transport gives us a path 
$\hat{V}(t)$ in $U(N_{B})$
\begin{eqnarray}
\hat{V}(t)={\bf{P}}\!\left[\exp\left(\int_{0}^{t}\hat{A}(s)ds\right)\right],
\end{eqnarray}
where ${\bf{P}}$ denotes path ordering. The holonomy for a closed path in $U(N_{A})$ is thus given by 
such an integral and is dependent on the Stokes matrix via the connection form in equation (\ref{gbnm}).

\label{exa}

\section{Relation to L\'{e}vay parallel transport for $SU(2)\times{SU}(2)$}
The pure two-qubit states can be represented as quaternionic qubit states \cite{mosseri01,levay04} using
the structure of the second Hopf-fibration. Within this representation one can construct the quaternionic
analogue of the Pancharatnam geometric phase, as has been done by L\'{e}vay \cite{levay04}. We review
this quaternionic representation and show that when the state evolution is generated by local $SU(2)$
operators, the L\'{e}vay geometric phase is contained in our construction.

In the quaternionic representation a pure two-qubit state

\begin{eqnarray}
\mid\!{\psi}\rangle=\alpha\!\mid\!{00}\rangle + \beta\!\mid\!{01}\rangle +
\gamma\!\mid\!{10}\rangle + \delta\!\mid\!{11}\rangle,
\end{eqnarray}
where $\alpha,\beta,\gamma,\delta\in{\mathbb{C}}$, is associated with a quaternionic qubit state

\begin{eqnarray}
\mid\!\mathbf{\Psi}\rangle = (\alpha+\beta{\mathbf{j}})\mid\!\mathbf{0}\rangle +
(\gamma+\delta{\mathbf{j}})\mid\!\mathbf{1}\rangle.
\end{eqnarray}
Here the quaternionic state $\mid\!\mathbf{0}$$\rangle$ is formed by identifying $\mid\!{00}\rangle$
and $\mid\!{01}\rangle$ in such a way that the complex coefficients of these states are contained in
the new quaterionic coefficient, and similarly for $\mid\!\mathbf{1}$$\rangle$. The standard quaternion
basis elements ${\mathbf{i}}$, ${\mathbf{j}}$ and ${\mathbf{k}}$, satisfy ${\mathbf{i}}^2 = {\mathbf{j}}^2 =
{\mathbf{k}}^2=-1$ and ${\mathbf{i}}{\mathbf{j}}=-{\mathbf{j}}{\mathbf{i}}={\mathbf{k}}$.
A local $SU(2)$ unitary $\hat{U}$ acting on the first qubit is represented by a operator
$\hat{\bf{U}}$ acting from the left on $\mid\!\mathbf{\Psi}\rangle$ and a local $SU(2)$
unitary $\hat{V}$ acting on the second qubit is represented by a unit quaternion $q_V$
acting from the right
\begin{eqnarray}
\mid\!\mathbf{\Psi}\rangle=\hat{{\bf{U}}}\mid\!\mathbf{\Psi_{0}}\rangle\hat{q}_V,
\end{eqnarray}
where $\hat{{\bf{U}}}={U}_{0}\hat{{\bf{1}}}+i\sum_{j=1}^{3}{U}_{j}\hat{{\boldsymbol{\sigma}}}_{j}$
with $\hat{{\bf{1}}}$ and $\hat{\boldsymbol{\sigma}}_{1},\hat{\boldsymbol{\sigma}}_{2},
\hat{\boldsymbol{\sigma}}_{3}$ the standard unit and Pauli operators acting on the quaternionic
qubit Hilbert space. The unit quaternion $\hat{q}_V\in{Sp(1)}$ corresponds to $\hat{V}$ according to
\begin{eqnarray}\label{asso}\hat{V}&=&V_{0}\hat{1}^{B}+i\sum_{j=1}^{3}V_{j}\hat{\sigma}_{j}^{B}
\nonumber\\
&&\phantom{hjhjh}\updownarrow\nonumber\\
\hat{q}_{V}&=&V_{0}+V_{3}\mathbf{i}-V_{2}\mathbf{j}+V_{1}\mathbf{k},
\end{eqnarray}
where $\hat{\sigma}_{1}^{B},\hat{\sigma}_{2}^{B},\hat{\sigma}_{3}^{B}$ are the standard Pauli
operators acting on the state space of the second qubit.

The inner product of two quaternionic states $\mid\!\!\mathbf{\Psi}\rangle =
p_{1}\!\mid\!\mathbf{0}\rangle+q_{1}\!\mid\!\mathbf{1}\rangle$ and $\mid\!\!\mathbf{\Phi}\rangle=p_{2}\!\mid\!\mathbf{0}\rangle+q_{2}\!\mid\!\mathbf{1}\rangle$ is
\begin{eqnarray}
\langle\mathbf{\Psi} \!\mid\!\mathbf{\Phi}\rangle=q_{1}^*q_{2}+p_{1}^*p_{2},
\end{eqnarray}
where $*$ is the quaternionic conjugation operation defined by $(a+b\mathbf{i} + c\mathbf{j} +
d\mathbf{k})^{\ast} = (a-b\mathbf{i}-c\mathbf{j}-d\mathbf{k})$, $a,b,c,d\in\mathbb{R}$.

\begin{figure}[h]
\includegraphics[width = 12 cm]{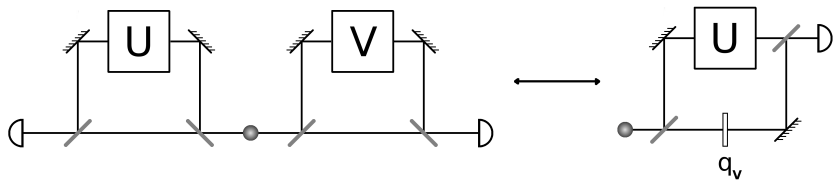}
\label{fig:fransmz}
\caption{\footnotesize{Correspondence between Franson setup and quaternionic QM Mach-Zehnder setup}}
\end{figure}

The quaternionic transition amplitude between two states related by a local $SU(2)$ operation
$\hat{U}$ on the first qubit can be expressed in terms of transition amplitudes in the ordinary
complex representation as
\begin{eqnarray}\label{assoo}\langle\mathbf{\Psi}\!\mid\hat{{\bf{U}}}\mid\!\mathbf{\Psi}\rangle = U_{0}+\sum_{j=1}^{3}U_{j}M_{j3}\mathbf{i}-\sum_{j=1}^{3}U_{j}M_{j2}\mathbf{j} +
\sum_{j=1}^{3}U_{j}M_{j1}\mathbf{k},
\end{eqnarray}
where $\mid\!\!\psi\rangle$ and $\mid\!\!\mathbf{\Psi}\rangle$ are the complex and quaternionic
representations of the same state. From this we can consider the formal analogue of the Mach-Zehnder
interference intensity
\begin{eqnarray}
I&=&\frac{1}{2}+\frac{1}{2}\re(\langle\mathbf{\Psi}\!\mid\hat{{\bf{U}}}\mid\!\mathbf{\Psi}\rangle\hat{q}_V)
\nonumber\\
&=&\frac{1}{2}+\frac{1}{2}U_{0}V_{0}-\frac{1}{2}\!\!\sum_{j,k=1}^{3}\!\!\!U_{j}V_{k}M_{jk},
\end{eqnarray}
where the phase factor $\hat{q}_V$ represents a local $SU(2)$ operation on the second qubit. To compare
this intensity to the Franson interference intensity for the case $\hat{U}\in{SU(2)}$ and $\hat{V}\in{SU(2)}$,
we consider equation (\ref{label}) when $D_{A}=2$ and use the parameterization $\hat{U}=U_{0}\hat{1}^{A} +
i\sum_{j=1}^{3}U_{j}\hat{\sigma}_{j}^{A}$, where $U_{0},U_{1},U_{2},U_{3}$ are real numbers. We then find
that the quaternionic Mach-Zehnder interference intensity $I$, is identical to the Franson intensity $I^{AB}$,
which demonstrates the correspondence between the quantum-mechanical Franson setup and the quaternionic
quantum-mechanics Mach-Zehnder setup as shown in figure \ref{fig:fransmz}.

This quaternionic representation of the two-qubit states can be considered a generalization of the ordinary
complex spinor representation of single qubits. In the ordinary single qubit representation, the Hilbert
space is the space of normalized complex spinors $S^3$ and the projective Hilbert space is the space of
spinors modulo a phase factor $S^3/U(1)={S}^3/S^1=S^2=\mathbb{C}P^1$, the complex projective space of
complex dimension one. In this quaternionic representation of two-qubit states, the Hilbert space is
the space of normalized quaternionic spinors $S^7$. The quaternionic projective Hilbert space is the
space of quaternionic spinors modulo a unit quaternion acting from the left, representing, as mentioned
above, a $SU(2)={S}^{3}$ rotation of the second qubit. This implies that the quaternionic projective
Hilbert space is $S^7/S^3=S^4=\mathbb{H}P^1$, the quaternionic projective space of quaternionic dimension
one, or real dimension $4$. While the ordinary single qubit representation corresponds to the first Hopf
fibration $S^1\hookrightarrow{S}^3\hookrightarrow{S}^2$, the quaternionic two qubit representation
corresponds to the second Hopf fibration $S^3\hookrightarrow{S}^7\hookrightarrow{S}^4$. This quaternionic
two-qubit representation is much similar to a qubit in quaternionic quantum mechanics \cite{adleranandan}
except that in this representation the absolute quaternionic phase corresponds to a rotation of the second
qubit and thus is a measurable quantity.

Within this representation of pure two qubit states, L\'{e}vay \cite{levay04} studied the
quaternionic analogue of the Pancharatnam parallel transport. L\'{e}vay's parallelity condition
and related parallel transport are defined such that two quaternionic states are parallel if
their inner product is a real and positive number. This condition is in concordance with the
Mach-Zehnder analogue picture since the maximal intensity is achieved when the unit quaternion
phase factor $\hat{q}_V$ is such that $\langle\mathbf{\Psi}\!\mid\!\hat{U}\!\mid\!
\mathbf{\Psi}\rangle\hat{q}_V$ is real and positive. We next consider a family of spinors
that are stepwise unitarily evolved from the same initial state $\mid\!\mathbf{\Psi}_{0}\rangle$
and demand that in each step the initial and final states are parallel. The transition amplitude
between two consecutive spinors in this family is
\begin{eqnarray}\langle\mathbf{\Psi}_{n-1}\!\mid\!\mathbf{\Psi}_{n}\rangle = \langle\mathbf{\Psi}_{n-1}\! \mid\hat{{\bf{U}}}^{(n)}\mid\!\mathbf{\Psi}_{n-1}\rangle\hat{q}_{V}^{(n)} = \lambda , \phantom{hhhhhh}
\lambda\in{\mathbb{R}_{+}},\!\!\!\!\!\!
\nonumber\\
\end{eqnarray}
where $\mid\!\mathbf{\Psi}_{n-1}\rangle=\!\hat{{\bf{U}}}\!(n-1)\mid\!\mathbf{\Psi}_{0}\rangle{\hat{q}_V}(n-1)$,
and we use the notation $\hat{{\bf{U}}}(n-1)=\hat{{\bf{U}}}^{(n-1)}\hat{{\bf{U}}}^{(n-2)}\dots{\hat{{\bf{U}}}^{(1)}}$
and $\hat{q}_{V}(n-1)=\hat{q}_{V}^{(1)}\hat{q}_{V}^{(2)}\dots{\hat{q}_{V}^{(n-1)}}$. Thus
\begin{eqnarray}
\langle\mathbf{\Psi}_{n-1}\!\mid\hat{{\bf{U}}}^{(n)}\mid\!\mathbf{\Psi}_{n-1}\rangle=\lambda{q}_{V}^{(n)\ast},
\end{eqnarray}
or by using equations (\ref{asso}) and (\ref{assoo})
\begin{eqnarray}
U^{(n)}_{0}\! & + & \! \sum_{j=1}^3 \! U^{(n)}_{j}\!M_{j3}\mathbf{i}-\!\! \sum_{j=1}^3\!U^{(n)}_{j}\!M_{j2}\mathbf{j}+\!\!\sum_{j=1}^3\!U^{(n)}_{j}\!M_{j1}
\mathbf{k} =
\nonumber\\
&=&\lambda(V^{(n)}_{0}-V^{(n)}_{3}\mathbf{i}+V^{(n)}_{2}\mathbf{j}-V^{(n)}_{1}\mathbf{k}).
\end{eqnarray}
Hence
\begin{eqnarray}\label{knott}V^{(n)}_{0}&=&\frac{1}{\lambda}U^{(n)}_{0},\nonumber\\
V^{(n)}_{j}&=&-\frac{1}{\lambda}\sum_{k=1}^3U^{(n)}_{k}M_{kj}.\label{uuuuu}\end{eqnarray}
The parameter $\lambda$ must be chosen to normalize $V^{(n)}$, hence $\lambda^2=U^{(n)2}_{0}+\sum_{i}(\sum_{j}U^{(n)}_{j}M_{ji})^2$.
If we compare this to equation (\ref{tgb}), when $D_{A}=2$ and again use the parameterization $\hat{U}=U_{0}\hat{1}^{A}+i\sum_{j=1}^3U_{j}\hat{\sigma}_{j}^{A}$,
where $U_{0},U_{1},U_{2},U_{3}$ are real numbers, we find that this parallelity condition
is the same as that in equation (\ref{uuuuu}).

To find the L\'{e}vay connection, we consider the infinitesimal limit in which the parallel transport
condition reads
\begin{eqnarray}
\langle{d}\mathbf{\Psi}\!\mid\!\mathbf{\Psi}\rangle=0,
\end{eqnarray}
where $\mid\!\mathbf{\Psi}\rangle$ is the instantaneous state. If we only allow changes generated
by the local unitaries  $\hat{{\bf U}}$ and $q_V$ we have
\begin{eqnarray}
\langle{d}\mathbf{\Psi}\!\mid{\mathbf{\Psi}}\rangle =
d{q_V}^{\ast} q_V +\langle d \mathbf{\Psi} \! \mid{\hat{{\bf U}}} d{\hat{{\bf U}}}^{\dagger}
\mid{\mathbf{\Psi}} \rangle.
\end{eqnarray}
Imposing the parallel transport condition $\langle{d}\mathbf{\Psi}\!\mid\!{\mathbf{\Psi}}\rangle=0$,
and using equations (\ref{asso}) and (\ref{assoo}) we find this to be equivalent to

\begin{eqnarray}
\label{grott}
(dVV^{\dagger})_{j}=-\sum_{k=1}^3(dUU^{\dagger})_{k}M_{kj}.
\end{eqnarray}
To compare this expression with the connection in our construction we consider equation (\ref{khuy})
for $D_{A}=D_{B}=2$, and note that for $SU(2)$ all symmetric structure constants  $d_{jkl}$ are zero.
This gives the $B$ matrix the following form
\begin{eqnarray}B=\left(\begin{array}{cccc}1 & S_{01} & S_{02} & S_{03}\\
S_{10} & 1 & 0 & 0\\
S_{20} & 0 & 1& 0\\
S_{30} & 0 & 0 & 1\end{array}\right).\end{eqnarray}
By taking into account that $(dUU^{\dagger})_{0}=0$ and $(dVV^{\dagger})_{0}=0$ for $SU(2)$, we see that
only the correlation matrix $M$ will be relevant to the relation between $\hat{dUU}^{\dagger}$ and
$\hat{dVV}^{\dagger}$. Since $B_{jk}=\delta_{jk}$ for $j,k\neq0$ it immediately follows that equation
(\ref{khuy}) reduces to equation (\ref{grott}).

\label{gnutt}

\section{Conclusion}
We have constructed a parallel transport procedure in the same spirit as that of Pancharatnam
\cite{panca,berry1,ramaseshan}, in the sense that it defines parallelity with reference to maximization
of an interferometric quantity. The interferometric quantity chosen in this case is the coincidence
intensity of a Franson type interferometer. The phase is taken to be the local degrees of freedom of
one of the subsystems.

Given two different two-partite states related by local unitary evolution of one of the subsystems,
the unitary operation that needs to be applied to the other subsystem to achieve parallelity, depends
on the correlation present in the full bipartite system. Generally phase unitaries that correspond
to different parallel transports do not commute, however, when the system is uncorrelated, only Abelian
phase factors need to be applied. Thus the holonomy group related to the parallel transport condition
is Abelian if the bipartite state is uncorrelated, and a non-Abelian holonomy group can be said to
be correlation induced.

The procedure is defined for arbitrary bipartite systems, pure as well as mixed. In the infintesimal
limit of the parallel transport, the connection form can be found as a closed expression for arbitrary
dimension. On the other hand, finding a closed expression for parallel transport when the steps are
finite appears to be non-trivial in the general $U(D)$ case. The procedure can be restricted to
subgroups of the full unitary groups. In the pure two-qubit case when only $SU(2)$ operators are
considered it has been shown that this procedure is related to L\'{e}vay parallel transport \cite{levay04}.
Therefore our construction opens up for experimental tests of the L\'{e}vay geometric phase in the
special case of local $SU(2)$ evolutions.

\section*{Acknowledgments}
M.E. acknowledges support from the Swedish Research Council (VR).
M.S.W. and E.S. acknowledges support from the National Research Foundation and the Ministry of
Education (Singapore). M.S.W. also acknowledges a Erwin Schr\"{o}dinger Junior Research Fellowship.

\end{document}